 \documentclass[12pt, draftclsnofoot, onecolumn]{IEEEtran} 
\usepackage[LGR,T1]{fontenc}
\usepackage[utf8]{inputenc}
\usepackage[cmex10]{amsmath}
\usepackage{amssymb}
\usepackage{amsfonts}
\usepackage{graphicx}
\usepackage{wasysym}
\usepackage[ruled]{algorithm2e}
\usepackage{textcomp}
\usepackage{xcolor}
\usepackage{hyperref}
\usepackage{cite}
\usepackage{array}
\usepackage{euscript}
\usepackage{bigstrut}
\usepackage{tabularx}
\usepackage{epstopdf}
\usepackage{float}
\usepackage{url}
\usepackage{stfloats}
\usepackage{verbatim}
\usepackage{balance}

\makeatletter

\ProvideTextCommand{\~}{LGR}[1]{\char126#1}

\newcommand{\lyxmathsym}[1]{\ifmmode\begingroup\def\b@ld{bold}
  \text{\ifx\math@version\b@ld\bfseries\fi#1}\endgroup\else#1\fi}

\begin{document}
\bstctlcite{BSTcontrol}

\title{Blockage-Aware and Shadowing-Aware RIS-Assisted Joint Communication and Positioning for Urban Non-Terrestrial Networks}
\author{Muhammad  Khalil, \textit{Member, IEEE},    \textit{Ke Wang, \textit{Senior Member, IEEE}},  and Jinho Choi, \textit{Fellow, IEEE} 
  
\thanks{The authors Khalil and Wang are with the School of Engineering, RMIT University, Melbourne, Australia. Emails:$\,$ muhammad.khalil@rmit.edu.au,  and  ke.wang@rmit.edu.au. The author Choi is with  the University of Adelaide, Adelaide, Australia. Email: jinho.choi@adelaide.edu.au}
}
\maketitle


\begin{abstract}
Reconfigurable intelligent surfaces (RISs) have recently been investigated for non-terrestrial networks (NTNs), primarily to improve satellite communication performance. However, RIS-assisted urban NTN designs that jointly support communication reliability and user positioning under blockage, while retaining low online complexity, remain limited. This paper proposes a blockage-aware and shadowing-aware RIS-assisted framework for joint communication and positioning in an urban low-Earth-orbit (LEO) satellite downlink. A terrestrial RIS is exploited not only to reinforce a blockage-sensitive satellite–user link but also to create an additional reflected path that improves delay-domain positioning observability. To capture this dual role, a reduced two-dimensional positioning model is developed using the direct-path delay and the RIS-assisted excess delay, and the resulting position error bound (PEB) is integrated with the received signal-to-noise ratio (SNR) into a unified communication–positioning utility. A blockage-aware three-mode operating policy is then introduced to adapt the RIS priority among communication-oriented, balanced, and positioning-oriented operations according to the direct-link condition. To improve robustness in realistic urban environments, spatially correlated RIS–user shadowing is tracked across coherence blocks using a state-space model and a scalar Kalman filter, and the filtered estimate is incorporated into a robust codebook-based RIS selection strategy with low online complexity. Numerical results show that the proposed framework achieves a controllable SNR–PEB tradeoff, reduces PEB relative to communication-only RIS selection while maintaining competitive SNR performance, improves codeword-selection stability under shadowing uncertainty, validates the intended three-mode blockage-aware RIS operation, and increases the joint success probability with RIS size and phase resolution, with diminishing returns at high hardware complexity. These results highlight the benefit of jointly optimizing communication reliability and positioning observability in RIS-assisted urban NTN design.
\end{abstract}


\section{Introduction}

Non-terrestrial networks (NTNs), especially those enabled by low-Earth-orbit (LEO) satellites, are becoming an important component of future wireless systems because they can provide wide-area coverage and service continuity where terrestrial infrastructure is limited or unreliable. Many NTN applications are also inherently location-dependent, so communication reliability alone is insufficient; accurate user positioning is equally important for services such as emergency response, location verification, and context-aware operation. In dense urban environments, however, these two objectives are difficult to achieve jointly because the direct satellite link may suffer severe blockage, while positioning accuracy is further degraded by unfavorable propagation geometry and limited measurement diversity \cite{Dureppagari2023}.

Reconfigurable intelligent surfaces (RISs) offer a promising way to enhance urban NTN performance by introducing a controllable reflected path between the transmitter and the receiver. Unlike passive reflectors, RISs can reconfigure the propagation environment with relatively low hardware complexity \cite{Basar2019}. In satellite communications, RIS-assisted designs have been shown to mitigate practical impairments and improve satellite-to-ground transmission through passive beamforming and multi-RIS deployment \cite{Khalil2025, Khalil2024}. However, the RIS configuration that is best for communication is not necessarily the one that is best for positioning, because positioning depends on the geometric information carried by the reflected path rather than on signal strength alone. As a result, RIS design in this setting involves an inherent communication--positioning tradeoff \cite{ Emenonye2024TWC}. This mismatch becomes more pronounced when the direct satellite path is weakened by blockage and the RIS--user hop is affected by shadowing uncertainty, which motivates RIS control strategies that explicitly balance both objectives rather than optimizing communication alone.


Existing RIS-related research most relevant to this work can be grouped into three main directions. First, RIS-assisted satellite communications have mainly focused on communication-centric objectives such as interference mitigation, signal enhancement, and energy-efficient beamforming. Recent IEEE studies have shown that RIS can improve satellite--terrestrial communication performance under practical impairments and dynamic LEO operation \cite{Khalil2026NOMA,Khalil2025,Khalil2026TD3,Khalil2024,Bariah2022}. However, these works are primarily communication-oriented and do not explicitly address positioning observability or delay-domain localization support.

Second, RIS-assisted localization research has shown that intelligent reflections can improve positioning accuracy by increasing geometric diversity and strengthening the Fisher information of the received observations. In particular, prior IEEE studies have derived position and orientation error bounds and clarified the roles of RIS geometry, phase design, and near-/far-field effects in localization performance \cite{Elzanaty2021, Emenonye2023,Ozturk2024}. Nevertheless, these studies mainly consider terrestrial wireless settings and do not explicitly capture NTN-specific blockage, degraded satellite visibility, or shadowing-aware RIS control.

Third, practical RIS operation has increasingly moved toward low-overhead control through finite codebooks, which reduce online optimization complexity and signaling overhead. Such codebook-based designs are attractive for scalable and implementable RIS control \cite{An2024}. However, existing codebook approaches are not tailored to the joint communication--positioning problem considered here and do not address the effect of uncertain and spatially correlated RIS--user shadowing on online codeword-selection stability.

Despite this progress, the literature still lacks a unified RIS control framework for urban NTN that jointly accounts for communication reliability, delay-domain positioning observability, blockage-aware objective adaptation, and uncertainty in the RIS--user terrestrial hop under practical low-complexity operation. This gap is especially important because the RIS configuration that is favorable for communication enhancement is not generally the one that provides the strongest positioning support, particularly when the direct satellite path is weakened by blockage and the RIS--user hop is affected by local shadowing uncertainty. This gap directly motivates the blockage-aware and shadowing-aware joint RIS design proposed in this paper.

To address this gap, we formulate a low-complexity, blockage-aware, and shadowing-aware RIS control framework for joint communication and positioning in urban NTN. The main contributions are summarized as follows.

\begin{itemize}
\item We formulate a joint communication–positioning RIS control problem for urban NTN, in which the RIS-reflected path is exploited not only to enhance downlink reliability but also to improve delay-domain positioning observability under blockage-sensitive satellite access.

\item We develop a reduced two-dimensional delay-domain positioning framework for RIS-assisted NTN, where the direct-path delay and the RIS-assisted excess delay are translated into a Fisher-information-based positioning metric, namely the position error bound (PEB), that can be directly integrated into RIS control.

\item We propose a blockage-aware and shadowing-aware low-complexity RIS controller that balances received SNR and positioning accuracy through a three-mode priority adaptation policy and stabilizes codebook selection through filtered shadowing tracking and robust uncertainty-aware evaluation.

\item We demonstrate numerically that the proposed design achieves a controllable communication–positioning tradeoff, improves positioning accuracy relative to communication-oriented RIS selection while maintaining competitive SNR performance, and yields more stable RIS codeword decisions under shadowing uncertainty.
\end{itemize}


The remainder of this paper is organized as follows. Section~II introduces the considered urban NTN system model, including the propagation geometry, the blockage-aware channel formulation, the reduced delay-domain positioning observations, and the shadowing-state model. Section~III presents the proposed blockage-aware and shadowing-aware RIS controller, including the joint communication–positioning utility, the three-mode operating policy, the robust codebook-based selection strategy, and the corresponding online operation. Section~IV reports the numerical results and analyzes the resulting communication, positioning, switching-stability, and joint success trends. Section~V concludes the paper.

\section{System Model}

\begin{figure}[!t]
\centering
\includegraphics[width=\columnwidth]{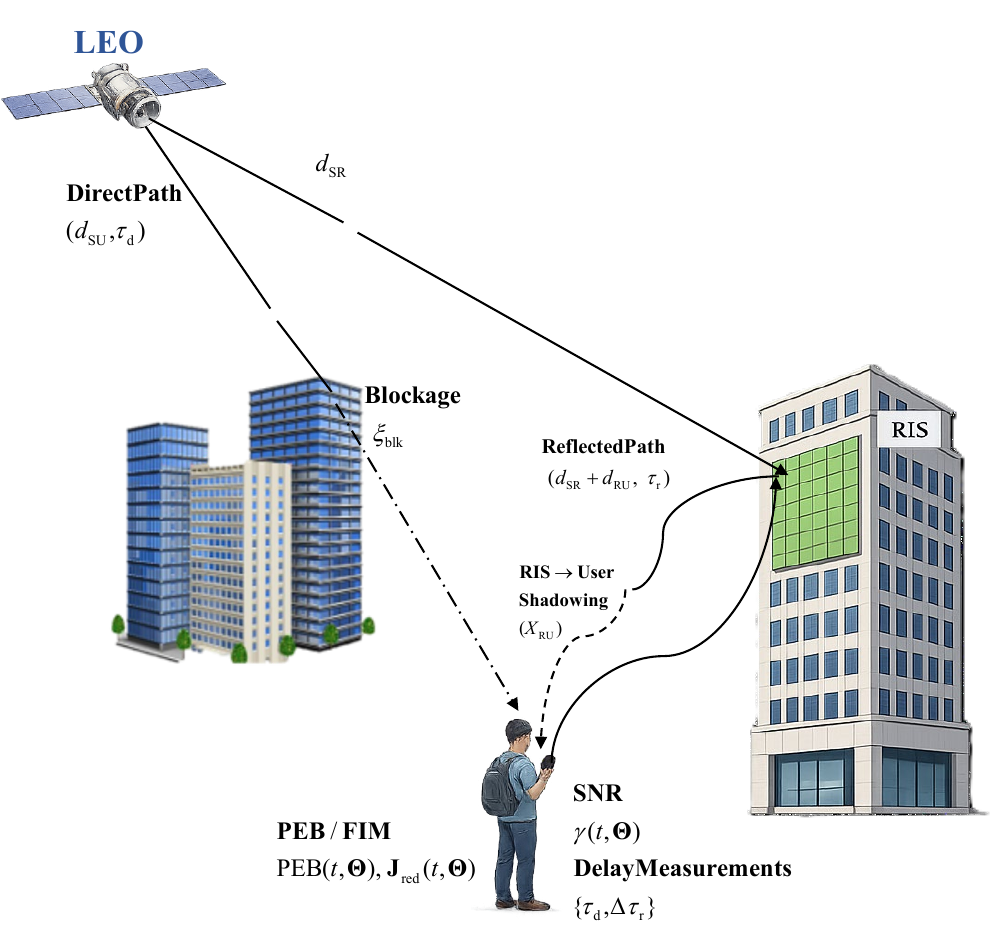}
\caption{System model of the considered RIS-assisted LEO NTN. A weak direct satellite--user path coexists with an RIS-assisted reflected path, enabling blockage-resilient communication and improved positioning observability in an urban environment.}
\label{fig:Fig_1}
\end{figure}

We consider a downlink NTN scenario in which a single LEO satellite serves a ground user in an urban environment with the assistance of a terrestrial RIS, as illustrated in Fig.~\ref{fig:Fig_1}. The proposed framework targets blockage-resilient operation in situations where the direct satellite-to-user link is weak or significantly attenuated. In such cases, the RIS establishes a controllable auxiliary propagation path that not only enhances the received signal quality but also improves the observability of the user position. In particular, the received signal consists of a weak direct component and an RIS-assisted reflected component, which together support joint communication and positioning in challenging urban settings.

\subsection{Network Geometry and Assumptions}

Let the satellite position at time $t$ and the fixed RIS position be denoted as follows:
\begin{equation}
\mathbf{p}_{\mathrm{S}}(t)\in\mathbb{R}^{3},
\qquad
\mathbf{p}_{\mathrm{R}}\in\mathbb{R}^{3},
\end{equation}
respectively, where the RIS location is assumed to be fixed and known, and the satellite ephemeris is assumed to be known from the network side. In addition, the RIS orientation is assumed to be fixed and known, such that the corresponding array responses are well defined.

For urban ground-user positioning, we adopt a reduced two-dimensional user-location model with a known height. Specifically, the user is assumed to move within a street-level service region, while the vertical coordinate is treated as known from the deployment context, e.g., a nominal pedestrian-height assumption consistent with standard urban channel-model settings or available map/terrain information \cite{TR38901}. Hence, the user position is expressed as
\begin{equation}
\mathbf{p}_{\mathrm{U}}(t)=
\begin{bmatrix}
x_{\mathrm{U}}(t)\\
y_{\mathrm{U}}(t)
\end{bmatrix}
\in\mathbb{R}^{2},
\qquad
z_{\mathrm{U}}=z_{0},
\label{eq:user_2d}
\end{equation}
where $z_{0}$ denotes the known street-level user height. For notational convenience, the corresponding three-dimensional user position is
\begin{equation}
\tilde{\mathbf{p}}_{\mathrm{U}}(t)=
\begin{bmatrix}
\mathbf{p}_{\mathrm{U}}^{\mathrm{T}}(t) & z_{0}
\end{bmatrix}^{\mathrm{T}}
\in\mathbb{R}^{3}.
\label{eq:user_3d_embed}
\end{equation}
This assumption is appropriate for the considered street-level urban scenario and enables a reduced horizontal positioning state; extension to unknown user height would require a full three-dimensional positioning model and is left for future work.

Accordingly, for the reduced delay-domain positioning analysis within one coherence block, the unknown user state is defined as
\begin{equation}
\mathbf{x}(t)=\mathbf{p}_{\mathrm{U}}(t)\in\mathbb{R}^{2}.
\label{eq:state_2d}
\end{equation}

Since the RIS is configured once per coherence block, we adopt a blockwise quasi-static geometry model. Specifically, the satellite ephemeris is assumed to be known from the network side, and the satellite–RIS–user geometry is treated as approximately constant within each coherence block. Thus, for a given block, the satellite position is evaluated at the corresponding time instant and regarded as fixed during that block, while it is updated from one block to the next according to the satellite motion.

The relevant propagation distances are
\begin{align}
d_{\mathrm{SU}}(t) &= \left\|\mathbf{p}_{\mathrm{S}}(t)-\tilde{\mathbf{p}}_{\mathrm{U}}(t)\right\|_{2}, \\
d_{\mathrm{SR}}(t) &= \left\|\mathbf{p}_{\mathrm{S}}(t)-\mathbf{p}_{\mathrm{R}}\right\|_{2}, \\
d_{\mathrm{RU}}(t) &= \left\|\mathbf{p}_{\mathrm{R}}-\tilde{\mathbf{p}}_{\mathrm{U}}(t)\right\|_{2}.
\end{align}

We consider a single-antenna user and an RIS composed of $N$ nearly passive reflecting elements. The RIS is configured once per coherence block through the following diagonal phase-shift matrix:
\begin{equation}
\mathbf{\Theta}(t)=
\mathrm{diag}
\left(
e^{j\phi_{1}(t)},e^{j\phi_{2}(t)},\ldots,e^{j\phi_{N}(t)}
\right),
\label{eq:Theta}
\end{equation}
where $\phi_{n}(t)\in\mathcal{F}_{b}$ is the phase of the $n$th RIS element, and
\begin{equation}
\mathcal{F}_{b}=
\left\{
0,\frac{2\pi}{2^{b}},\ldots,\frac{2\pi(2^{b}-1)}{2^{b}}
\right\}
\end{equation}
denotes the feasible set for a $b$-bit phase shifter.

\subsection{Blockage-Aware Channel Model}

For the direct satellite-to-user link, we separate the nominal large-scale attenuation from the additional urban blockage factor, following the blockage-sensitive NTN modeling philosophy adopted in 3GPP-oriented NTN studies \cite{TR38811}:
\begin{equation}
y(t)=
\left(
h_{\mathrm{d}}(t)
+
h_{\mathrm{r}}(t,\boldsymbol{\Theta}(t))
\right)s(t)+w(t),
\label{eq:rx_signal}
\end{equation}
where $s(t)$ is the transmitted pilot or data symbol with average power $P_{\mathrm{t}}$, $w(t)\sim\mathcal{CN}(0,\sigma_{w}^{2})$ is additive white Gaussian noise, $h_{\mathrm{d}}(t)$ is the direct satellite-to-user channel, and $h_{\mathrm{r}}(t,\mathbf{\Theta})$ is the RIS-assisted reflected channel.

The direct channel is modeled as
\begin{equation}
h_{\mathrm{d}}(t)=
\sqrt{\beta_{\mathrm{SU}}(t)}\,
e^{-j2\pi f_{\mathrm{c}}\tau_{\mathrm{d}}(t)},
\label{eq:hd}
\end{equation}
where $f_{\mathrm{c}}$ is the carrier frequency, and $\tau_{\mathrm{d}}(t)$ is the direct-path delay.

The RIS-assisted channel is represented as follows:
\begin{equation}
h_{\mathrm{r}}(t,\mathbf{\Theta})
=
\mathbf{h}_{\mathrm{RU}}^{\mathrm{T}}(t)\,
\mathbf{\Theta}(t)\,
\mathbf{h}_{\mathrm{SR}}(t),
\label{eq:hr}
\end{equation}
where $\mathbf{h}_{\mathrm{SR}}(t)\in\mathbb{C}^{N\times 1}$ and $\mathbf{h}_{\mathrm{RU}}(t)\in\mathbb{C}^{N\times 1}$ denote the satellite-to-RIS and RIS-to-user channels, respectively. 

To capture the dominant controllable RIS-assisted component, we adopt a dominant line-of-sight (LoS) or specular model for the satellite-to-RIS and RIS-to-user channels. This approximation is particularly appropriate for the satellite-to-RIS link and for the dominant reflected component shaped by the RIS phase profile. In the considered urban street scenario, finer diffuse multipath effects on the terrestrial RIS-to-user hop are not modeled explicitly at the small-scale; instead, their average impact is absorbed into the corresponding large-scale attenuation and shadowing terms. Under this model, the channels are expressed as
\begin{align}
\mathbf{h}_{\mathrm{SR}}(t)
&=
\sqrt{\beta_{\mathrm{SR}}(t)}\,
\mathbf{a}_{\mathrm{SR}}(t)\,
e^{-j2\pi f_{\mathrm{c}}\tau_{\mathrm{SR}}(t)},
\\
\mathbf{h}_{\mathrm{RU}}(t)
&=
\sqrt{\beta_{\mathrm{RU}}(t)}\,
\mathbf{a}_{\mathrm{RU}}(t)\,
e^{-j2\pi f_{\mathrm{c}}\tau_{\mathrm{RU}}(t)},
\end{align}
where $\mathbf{a}_{\mathrm{SR}}(t)$ and $\mathbf{a}_{\mathrm{RU}}(t)$ denote the RIS array response vectors, and
\begin{equation}
\tau_{\mathrm{SR}}(t)=\frac{d_{\mathrm{SR}}(t)}{c},
\qquad
\tau_{\mathrm{RU}}(t)=\frac{d_{\mathrm{RU}}(t)}{c},
\end{equation}
with $c$ denoting the speed of light.

The large-scale gain of each link is modeled in linear scale.
For the direct satellite-to-user link, we separate the nominal
large-scale attenuation from the additional urban blockage factor:
\begin{equation}
\beta_{\mathrm{SU}}(t)=\xi_{\mathrm{blk}}(t)\,\bar{\beta}_{\mathrm{SU}}(t),
\qquad
\xi_{\mathrm{blk}}(t)\in(0,1],
\label{eq:blockage_factor}
\end{equation}
where $\xi_{\mathrm{blk}}(t)=1$ corresponds to an unblocked direct path,
whereas $\xi_{\mathrm{blk}}(t)\ll 1$ represents severe attenuation due to
blockage. In this work, we restrict attention to the regime
$\xi_{\mathrm{blk}}(t)>0$, such that a weak but detectable direct timing
reference remains available.

The nominal direct-link gain is
\begin{equation}
\bar{\beta}_{\mathrm{SU}}(t)=10^{-\bar{L}_{\mathrm{SU}}(t)/10},
\label{eq:beta_su_nominal}
\end{equation}
with
\begin{equation}
\bar{L}_{\mathrm{SU}}(t)=
L_{\mathrm{FS}}\!\big(d_{\mathrm{SU}}(t),f_{\mathrm{c}}\big)
+
L_{\mathrm{atm},\mathrm{SU}}(t)
+
X_{\mathrm{SU}}(t),
\label{eq:Lsu_nominal}
\end{equation}
where $L_{\mathrm{atm},\mathrm{SU}}(t)$ denotes the atmospheric attenuation
on the direct satellite--user link in dB, and $X_{\mathrm{SU}}(t)$ denotes
an additional large-scale excess-loss term in dB, capturing slow
environmental attenuation on the direct link beyond free-space and
atmospheric losses. Since the explicit blockage effect is modeled
separately through $\xi_{\mathrm{blk}}(t)$, $X_{\mathrm{SU}}(t)$ does not represent
the blockage-sensitive attenuation captured by $\xi_{\mathrm{blk}}(t)$.

Moreover,
\begin{equation}
L_{\mathrm{FS}}(d,f_{\mathrm{c}})
=
20\log_{10}\!\left(\frac{4\pi f_{\mathrm{c}}d}{c}\right).
\label{eq:FSPL}
\end{equation}
For the satellite-to-RIS link, the large-scale gain is expressed as
\begin{equation}
\beta_{\mathrm{SR}}(t)=10^{-L_{\mathrm{SR}}(t)/10},
\label{eq:beta_sr}
\end{equation}
where
\begin{equation}
L_{\mathrm{SR}}(t)=
L_{\mathrm{FS}}(d_{\mathrm{SR}}(t),f_{\mathrm{c}})
+
L_{\mathrm{atm},\mathrm{SR}}(t)
+
X_{\mathrm{SR}}(t).
\label{eq:Lsr}
\end{equation}
For the terrestrial RIS-to-user hop, we adopt
\begin{equation}
\beta_{\mathrm{RU}}(t)=10^{-L_{\mathrm{RU}}(t)/10},
\label{eq:beta_ru}
\end{equation}
with
\begin{equation}
L_{\mathrm{RU}}(t)=
L_{\mathrm{FS}}(d_{\mathrm{RU}}(t),f_{\mathrm{c}})
+
X_{\mathrm{RU}}(t),
\label{eq:Lru}
\end{equation}
where $X_{\mathrm{RU}}(t)$ denotes the large-scale shadowing term on the terrestrial RIS-to-user hop. Unlike the direct satellite-to-user link, whose additional attenuation is explicitly modeled by the blockage factor $\xi_{\mathrm{blk}}(t)$, the reflected RIS-assisted path is affected through the local RIS-to-user propagation condition embedded in $X_{\mathrm{RU}}(t)$. Hence, partial obstruction and local clutter effects on the RIS-assisted path are modeled implicitly through the RIS-to-user shadowing process rather than through a separate binary or multiplicative blockage factor. This is consistent with the considered urban street scenario, where the terrestrial reflected hop is mainly governed by near-ground propagation and local environmental shadowing. A separate explicit blockage state for complete obstruction of the RIS-assisted path is not introduced here; such an extension could be incorporated in future work by augmenting the reflected-hop model with an additional blockage factor. For online RIS control, the local shadowing term on the RIS-to-user hop is further tracked across coherence blocks using the state-space model introduced in Section~\ref{sec:spatially_correlated_shadowing}.

\subsection{Spatially Correlated Shadowing}
\label{sec:spatially_correlated_shadowing}

Because the user moves within a cluttered urban environment, the shadowing on the terrestrial RIS-to-user hop is spatially correlated. To model this effect, the shadowing field is represented as a Gaussian random field with exponential covariance
\begin{equation}
\mathbb{E}
\left[
X_{\mathrm{RU}}(\mathbf{p})\,X_{\mathrm{RU}}(\mathbf{p}')
\right]
=
\sigma_{\mathrm{RU}}^{2}
\exp\!\left(
-\frac{\|\mathbf{p}-\mathbf{p}'\|_{2}}{d_{\mathrm{corr}}}
\right),
\label{eq:shadow_cov}
\end{equation}
where $\sigma_{\mathrm{RU}}^{2}$ denotes the variance of the large-scale shadowing on the terrestrial RIS-to-user hop, reflecting the severity of urban-clutter-induced attenuation variations, and $d_{\mathrm{corr}}$ denotes the corresponding correlation distance, i.e., the spatial scale over which nearby street-level user locations experience similar local propagation conditions. This model captures the fact that nearby user positions experience similar local propagation conditions, which are important for both outage analysis and tracking stability. Such spatially correlated shadowing behavior is also consistent with established wireless channel modeling practices and motivates the use of recursive shadowing-state estimation in mobile environments \cite{TR38811,Jiang2003}.
\subsection{Shadowing-State Tracking for Online RIS Control}
\label{sec:shadow_tracking}

While \eqref{eq:shadow_cov} models the spatial correlation of the RIS–user shadowing field over space, the RIS controller can further exploit this structure in online operations by tracking the local shadowing state across coherence blocks. To this end, we introduce a discrete-time block index $k$.

Let $X_{\mathrm{RU}}[k]$ denote the local RIS–user shadowing component in dB at block $k$. Motivated by the exponential spatial correlation model in \eqref{eq:shadow_cov}, we adopt the first-order state evolution, which is consistent with the autoregressive shadowing and local-mean power tracking models used in wireless Kalman-filter-based estimation \cite{Jiang2003}.
\begin{equation}
X_{\mathrm{RU}}[k]
=
\rho[k]\,X_{\mathrm{RU}}[k-1] + v[k],
\label{eq:shadow_state}
\end{equation}
where $v[k]\sim\mathcal{N}(0,q_X)$ is the process noise, and
\begin{equation}
\rho[k]
=
\exp\!\left(-\frac{\Delta s[k]}{d_{\mathrm{corr}}}\right)
\label{eq:shadow_rho}
\end{equation}
captures the spatial memory induced by user mobility, with
\begin{equation}
\Delta s[k]
=
\left\|
\mathbf{p}_{\mathrm{U}}(t_k)-\mathbf{p}_{\mathrm{U}}(t_{k-1})
\right\|_2
\end{equation}
denoting the user displacement between two consecutive coherence blocks.

To obtain an online observation of the shadowing state, we use the reflected pilot-power measurement after removing the deterministic geometry-dependent component. Specifically, define
\begin{equation}
y_{\mathrm{sh}}[k]
=
\bar{P}_{\mathrm{r,dB}}[k]
-
P_{\mathrm{r,dB}}^{\mathrm{meas}}[k]
=
X_{\mathrm{RU}}[k] + n_X[k],
\label{eq:shadow_meas}
\end{equation}
where $\bar{P}_{\mathrm{r,dB}}[k]$ denotes the nominal reflected pilot power predicted at block $k$ from the known geometry and the large-scale path-loss model, excluding the local RIS–user shadowing term; $P_{\mathrm{r,dB}}^{\mathrm{meas}}[k]$ denotes the corresponding measured reflected pilot power; and $n_X[k]\sim\mathcal{N}(0,r_X)$ denotes the effective observation noise. In practice, the predicted quantity $\bar{P}_{\mathrm{r,dB}}[k]$ may be imperfect due to residual geometry mismatch, uncertainty in large-scale path-loss parameters, or hardware and measurement imperfections. Such prediction mismatch is absorbed into the effective observation uncertainty and is handled through the measurement-noise variance $r_X$, while the process-noise variance $q_X$ captures the actual block-to-block evolution of the local shadowing state.

A scalar Kalman filter is then applied to obtain the filtered shadowing estimate $\hat{X}_{RU}[k|k]$ and its estimation error variance $P_X[k|k]$. Here, $q_X$ and $r_X$ denote the Kalman process-noise and measurement-noise variances, respectively. The parameter $q_X$ controls the allowed block-to-block variability of the local RIS–user shadowing state beyond the spatial memory captured by $\rho[k]$, whereas $r_X$ reflects the uncertainty of the shadowing observation in \eqref{eq:shadow_meas}, including reflected pilot-power measurement noise and residual mismatch in the nominal predicted power. The corresponding prediction and update steps are
\begin{align}
\hat{X}_{\mathrm{RU}}[k|k-1]
&=
\rho[k]\hat{X}_{\mathrm{RU}}[k-1|k-1],
\\
P_X[k|k-1]
&=
\rho^2[k]P_X[k-1|k-1] + q_X,
\\
K_X[k]
&=
\frac{P_X[k|k-1]}{P_X[k|k-1]+r_X},
\\
\hat{X}_{\mathrm{RU}}[k|k]
&=
\hat{X}_{\mathrm{RU}}[k|k-1] \notag\\
&\quad+
K_X[k]\bigl(y_{\mathrm{sh}}[k]-\hat{X}_{\mathrm{RU}}[k|k-1]\bigr),
\\
P_X[k|k]
&=
\bigl(1-K_X[k]\bigr)P_X[k|k-1].
\label{eq:shadow_kf}
\end{align}
In practice, these parameters can be selected from prior channel statistics or tuned from pilot data so that the innovation sequence remains consistent with the assumed model. In the numerical evaluation, they are treated as filter-design parameters chosen to provide stable shadowing tracking under the considered urban mobility and measurement conditions. The pair $\bigl(\hat{X}_{\mathrm{RU}}[k|k],P_X[k|k]\bigr)$ therefore provides the controller with both a filtered estimate of the local shadowing condition and a measure of its uncertainty, which will be used in the robust RIS codebook selection developed in the next section.

\subsection{Communication and Positioning Observations}

For communication, the downlink reliability is characterized by the instantaneous SNR
\begin{equation}
\gamma(t,\mathbf{\Theta})=
\frac{P_{\mathrm{t}}}{\sigma_{w}^{2}}
\left|
h_{\mathrm{d}}(t)+h_{\mathrm{r}}(t,\mathbf{\Theta})
\right|^{2}.
\label{eq:gamma}
\end{equation}

For positioning, the satellite is assumed to transmit known wideband pilot signals that allow the receiver to estimate the direct-path delay and the RIS-assisted excess delay. This delay-domain positioning viewpoint is consistent with RIS-assisted positioning studies that exploit additional reflected-path observations to strengthen the Fisher information of the received measurements. Hence, the communication and positioning layers use different but mutually consistent signal abstractions.

To obtain the reduced positioning model considered in this work, we assume that the user has already acquired the NTN downlink timing reference, so that absolute direct-path delay measurements are available after receiver timing acquisition. This assumption is adopted to isolate the communication–positioning role of the RIS in a blockage-sensitive urban scenario. Without such timing acquisition, an additional receiver clock-bias parameter would need to be introduced, and the problem would become a joint synchronization-and-positioning formulation, which is beyond the scope of the present paper. Under this assumption, the direct-path propagation delay is
\begin{equation}
\tau_{\mathrm{d}}(t)=\frac{d_{\mathrm{SU}}(t)}{c},
\label{eq:tau_d}
\end{equation}
while the RIS-assisted path delay is
\begin{equation}
\tau_{\mathrm{r}}(t)=\frac{d_{\mathrm{SR}}(t)+d_{\mathrm{RU}}(t)}{c}.
\label{eq:tau_r}
\end{equation}
Therefore, the RIS excess delay relative to the direct path is
\begin{equation}
\Delta\tau_{\mathrm{r}}(t)=\tau_{\mathrm{r}}(t)-\tau_{\mathrm{d}}(t)
=
\frac{d_{\mathrm{SR}}(t)+d_{\mathrm{RU}}(t)-d_{\mathrm{SU}}(t)}{c}.
\label{eq:delay_excess}
\end{equation}

For the numerical study, we adopt a reduced two-dimensional positioning model with a known user height. Accordingly, the positioning analysis is based on the reduced delay-domain observation subset
\begin{equation}
\hat{\mathbf{z}}_{\mathrm{red}}(t)=
\begin{bmatrix}
\hat{\tau}_{\mathrm{d}}(t)\\
\widehat{\Delta\tau}_{\mathrm{r}}(t)
\end{bmatrix}
=
\mathbf{h}_{\mathrm{red}}\!\left(\mathbf{p}_{\mathrm{U}}(t)\right)
+
\mathbf{n}_{\mathrm{red}}(t),
\label{eq:obs_model_reduced}
\end{equation}
where
\begin{equation}
\mathbf{h}_{\mathrm{red}}\!\left(\mathbf{p}_{\mathrm{U}}(t)\right)
=
\begin{bmatrix}
\tau_{\mathrm{d}}(t)\\
\Delta\tau_{\mathrm{r}}(t)
\end{bmatrix}.
\end{equation}

The reduced observation noise is modeled as follows:
\begin{equation}
\mathbf{n}_{\mathrm{red}}(t)\sim
\mathcal{N}\!\left(
\mathbf{0},
\mathbf{R}_{\mathrm{red}}(t,\mathbf{\Theta})
\right),
\end{equation}
with the covariance matrix
\begin{equation}
\mathbf{R}_{\mathrm{red}}(t,\mathbf{\Theta})
=
\begin{bmatrix}
\sigma_{\tau_{\mathrm{d}}}^{2}(t) & 0\\
0 & \sigma_{\Delta\tau_{\mathrm{r}}}^{2}(t,\mathbf{\Theta})
\end{bmatrix},
\label{eq:Rred}
\end{equation}
where, for simplicity, the direct-delay and excess-delay estimation errors are assumed to be uncorrelated. Although the geometric quantities $\tau_{\mathrm{d}}(t)$ and $\Delta\tau_{\mathrm{r}}(t)$ are determined by the propagation paths, their estimation accuracy depends on the quality of the received signal. Motivated by the well-known Cram\'er--Rao lower bound (CRLB) scaling of time-delay estimation with received SNR and effective bandwidth \cite{Shen2010}, we adopt the following simplified delay estimation variance models:
\begin{equation}
\sigma_{\tau_{\mathrm{d}}}^{2}(t)
=
\frac{\kappa_{\mathrm{d}}}{B^{2}\,\gamma_{\mathrm{d}}(t)},
\label{eq:tau_d_var_model}
\end{equation}
and
\begin{equation}
\sigma_{\Delta\tau_{\mathrm{r}}}^{2}(t,\mathbf{\Theta})
=
\frac{\kappa_{\mathrm{r}}}{B^{2}\,\gamma_{\mathrm{r}}(t,\mathbf{\Theta})},
\label{eq:tau_r_var_model}
\end{equation}
where $B$ denotes the nominal signal bandwidth, used here as a proxy for the effective bandwidth, and $\kappa_{\mathrm{d}}>0$ and $\kappa_{\mathrm{r}}>0$ are constants that absorb waveform-dependent factors and estimator implementation effects for the direct and reflected-path delay measurements, respectively. Furthermore,
\begin{equation}
\gamma_{\mathrm{d}}(t)
=
\frac{P_{\mathrm{t}}}{\sigma_{w}^{2}}
\left|h_{\mathrm{d}}(t)\right|^{2}
\label{eq:gamma_d}
\end{equation}
is the direct-path estimation SNR, while
\begin{equation}
\gamma_{\mathrm{r}}(t,\mathbf{\Theta})
=
\frac{P_{\mathrm{t}}}{\sigma_{w}^{2}}
\left|h_{\mathrm{r}}(t,\mathbf{\Theta})\right|^{2}
\label{eq:gamma_r}
\end{equation}
is the reflected-path estimation SNR. These models preserve the key first-order dependence that the delay estimation becomes more accurate as the received SNR and signal bandwidth increase.

The resulting equivalent Fisher information matrix for the two-dimensional user position is defined as
\begin{equation}
\mathbf{J}_{\mathrm{red}}(t,\mathbf{\Theta})
=
\left(
\frac{\partial \mathbf{h}_{\mathrm{red}}}{\partial \mathbf{p}_{\mathrm{U}}}
\right)^{\mathrm{T}}
\mathbf{R}_{\mathrm{red}}^{-1}(t,\mathbf{\Theta})
\left(
\frac{\partial \mathbf{h}_{\mathrm{red}}}{\partial \mathbf{p}_{\mathrm{U}}}
\right),
\label{eq:FIM_reduced}
\end{equation}
where
\begin{equation}
\frac{\partial \mathbf{h}_{\mathrm{red}}}{\partial \mathbf{p}_{\mathrm{U}}}
=
\begin{bmatrix}
\frac{\partial \tau_{\mathrm{d}}}{\partial x_{\mathrm{U}}} &
\frac{\partial \tau_{\mathrm{d}}}{\partial y_{\mathrm{U}}}
\\[1mm]
\frac{\partial \Delta\tau_{\mathrm{r}}}{\partial x_{\mathrm{U}}} &
\frac{\partial \Delta\tau_{\mathrm{r}}}{\partial y_{\mathrm{U}}}
\end{bmatrix}.
\end{equation}
The corresponding position error bound is
\begin{equation}
\mathrm{PEB}(t,\mathbf{\Theta})
=
\sqrt{
\mathrm{tr}
\left(
\mathbf{J}_{\mathrm{red}}^{-1}(t,\mathbf{\Theta})
\right)
}.
\label{eq:PEB}
\end{equation}
This reduced formulation preserves the key geometric contribution of the RIS-assisted excess-delay measurement and yields a locally identifiable two-dimensional positioning model under non-degenerate satellite--RIS--user geometry.

The derivatives in \eqref{eq:FIM_reduced} are taken with respect to the two-dimensional user coordinates, while the propagation distances are evaluated through the embedded three-dimensional position in \eqref{eq:user_3d_embed} at a fixed height of $z_{0}$.

Therefore, the RIS configuration affects both communication and positioning performance through a pair of metrics
\begin{equation}
\left\{
\gamma(t,\mathbf{\Theta}),\,
\mathrm{PEB}(t,\mathbf{\Theta})
\right\},
\end{equation}
which forms the basis for the joint RIS design developed in the next section.

\section{Joint RIS Design for Communication--Positioning Resilience}

Building on the system model in Section II, the RIS must be configured to improve not only the communication reliability of the satellite downlink but also the positioning observability of the user under urban blockage. Since these two objectives are generally coupled and may not be simultaneously maximized by the same RIS configuration, we adopt a joint communication–positioning design framework. In addition, because the terrestrial RIS-user hop may experience locally severe and spatially correlated shadowing, the controller is further enhanced with a shadowing-aware robust selection layer. As a result, the proposed design is both blockage-aware and shadowing-aware.

\subsection{Joint Design Objective}

For a given coherence block, let the RIS configuration be represented by $\mathbf{\Theta}(t)$. From Section II, the communication quality is characterized by the instantaneous SNR in \eqref{eq:gamma}, while the positioning performance is measured through the PEB in \eqref{eq:PEB}.

Because $\gamma(t,\Theta)$ should be maximized whereas $\mathrm{PEB}(t,\Theta)$ should be minimized, we define the following normalized nominal utility:
\begin{equation}
\mathcal{U}(t,\mathbf{\Theta})
=
\alpha(t)\,
\frac{\gamma(t,\mathbf{\Theta})}{\gamma_{\mathrm{ref}}}
+
\bigl(1-\alpha(t)\bigr)\,
\frac{\mathrm{PEB}_{\mathrm{ref}}}{\mathrm{PEB}(t,\mathbf{\Theta})},
\label{eq:utility}
\end{equation}
where $\gamma_{\mathrm{ref}}>0$ and $\mathrm{PEB}_{\mathrm{ref}}>0$ are fixed normalization constants used to place the communication and positioning terms on comparable dimensionless scales. In practice, these reference values are selected offline from representative nominal SNR and PEB levels for the considered urban NTN scenario, for example, from a baseline RIS configuration or a reference operating point under typical geometry and blockage conditions. They are not optimized online and are kept fixed across coherence blocks and candidate RIS configurations so that the utility provides a consistent communication–positioning tradeoff metric. In addition, $\alpha(t)$ controls the relative importance of communication and positioning. A larger $\alpha(t)$ emphasizes communication reliability, whereas a smaller $\alpha(t)$ prioritizes positioning accuracy.

Accordingly, the nominal RIS design problem is formulated as
\begin{align}
\max_{\mathbf{\Theta}(t)} \quad
& \mathcal{U}(t,\mathbf{\Theta})
\label{eq:opt_main}
\\
\text{s.t.}\quad
& \phi_n(t)\in\mathcal{F}_{b}, \qquad n=1,\ldots,N.
\nonumber
\end{align}

The formulation in \eqref{eq:opt_main} directly captures the dual role of the RIS. The first term in \eqref{eq:utility} promotes coherent signal enhancement through the reflected path, while the second term rewards RIS configurations that improve reflected path positioning support by reducing the PEB. As shown later, the actual proposed controller evaluates this utility in a robust manner by accounting for the filtered estimate of the local RIS–user shadowing condition.

\subsection{Blockage-Aware Operating Modes}

In urban NTN operation, the relative importance of communication and positioning varies with the direct-path condition. When the direct satellite path is healthy, the RIS mainly acts as a communication enhancer. In contrast, under severe blockage, the RIS should also improve the observability of the user position through the additional reflected-path measurement. To capture this effect, $\alpha(t)$ can be adaptively decided.

Let $\xi_{\mathrm{blk}}(t)\in(0,1]$ denote the direct-link blockage factor defined in \eqref{eq:blockage_factor}. We consider three operating modes:
\begin{equation}
\alpha(t)=
\begin{cases}
\alpha_{\mathrm{C}}, & \xi_{\mathrm{blk}}(t)\geq \xi_{\mathrm{H}},
\\[1mm]
\alpha_{\mathrm{B}}, & \xi_{\mathrm{L}}<\xi_{\mathrm{blk}}(t)<\xi_{\mathrm{H}},
\\[1mm]
\alpha_{\mathrm{P}}, & \xi_{\mathrm{blk}}(t)\leq \xi_{\mathrm{L}},
\end{cases}
\label{eq:alpha_modes}
\end{equation}
where $\xi_{\mathrm{H}}$ and $\xi_{\mathrm{L}}$ are two blockage thresholds that satisfy $0<\xi_{\mathrm{L}}<\xi_{\mathrm{H}}\leq1$, and
\begin{equation}
1 \ge \alpha_{\mathrm{C}} > \alpha_{\mathrm{B}} > \alpha_{\mathrm{P}} \ge 0.
\end{equation}
Here:
\begin{itemize}
\item $\alpha_{\mathrm{C}}$ corresponds to a \emph{communication priority mode};
\item $\alpha_{\mathrm{B}}$ corresponds to a \emph{balanced mode};
\item $\alpha_{\mathrm{P}}$ corresponds to a \emph{positioning support mode}.
\end{itemize}
The thresholds $(\xi_{\mathrm{L}},\xi_{\mathrm{H}})$ and the mode weights $(\alpha_{\mathrm{C}},\alpha_{\mathrm{B}},\alpha_{\mathrm{P}})$ are treated as offline design parameters that reflect the desired system policy for balancing communication reliability and positioning accuracy under different blockage conditions. In practice, they can be calibrated from application-level requirements such as target communication reliability and positioning accuracy or selected through offline sensitivity studies and a grid search over representative blockage regimes and user geometries. Once chosen, these parameters are kept fixed during online operation, while the mode itself is selected adaptively through $\xi_{\mathrm{blk}}(t)$ according to \eqref{eq:alpha_modes}. In the numerical evaluation, the specific values reported in Table~\ref{tab:sim_parameters} are selected to provide a clear separation among communication-priority, balanced, and positioning-support operations in the considered urban NTN scenario.

This mode selection policy is physically intuitive. When the direct path is sufficiently strong, the communication objective dominates, and the RIS is primarily tuned to improve SNR. Under moderate blockage, a balanced operation is preferred. Under severe blockage, the direct path weakens, and the RIS is configured to preserve both decodability and geometric diversity, thereby improving blockage resilience.

Importantly, these three operating modes determine the communication--positioning priority through $\alpha(t)$, but they do not yet account for the uncertainty of the local RIS--user shadowing. This additional robustness aspect is incorporated next.

\subsection{Shadowing-Aware Robust RIS Selection}
\label{sec:shadow_robust}

Although the blockage-aware policy in \eqref{eq:alpha_modes} determines the desired communication–positioning priority based on the direct-link condition, the RIS–user hop may still experience severe and uncertain shadowing locally. If the RIS selection is based only on instantaneous or nominal quantities, the chosen codeword may become unstable in spatially correlated urban environments. To improve resilience, we augment the controller with a shadowing-aware robust selection layer based on the filtered shadowing estimate obtained from the state-space tracker in Section~\ref{sec:shadow_tracking}. Hereafter, the discrete-time notation $[k]$ is used for online RIS selection over coherence blocks, with $t_k$ denoting the time instant of block $k$ and $\alpha[k]\triangleq\alpha(t_k)$.

Specifically, let $\hat{X}_{\mathrm{RU}}[k|k]$ and $P_X[k|k]$ denote the filtered estimates of the RIS–user shadowing in dB and its estimation error variance at coherence block $k$, respectively. A conservative shadowing level is then defined as
\begin{equation}
\hat{X}_{\mathrm{RU}}^{\mathrm{wc}}[k]
=
\hat{X}_{\mathrm{RU}}[k|k]
+
\nu \sqrt{P_X[k|k]},
\label{eq:shadow_wc}
\end{equation}
where $\nu \ge 0$ is a robustness parameter. This construction provides a risk-aware conservative estimate of the RIS–user shadowing by augmenting the filtered mean shadowing level with a margin proportional to its estimation uncertainty. Since larger shadowing values correspond to larger path loss on the terrestrial RIS-to-user hop, the quantity $\hat{X}_{\mathrm{RU}}^{\mathrm{wc}}[k]$ represents a pessimistic operating point for robust RIS evaluation. In this sense, $\nu$ controls the conservativeness of the controller: larger values of $\nu$ place more weight on adverse shadowing realizations, whereas $\nu=0$ recovers the nominal estimate without an additional robustness margin.

Using \eqref{eq:shadow_wc}, the path loss for the RIS user used in codeword evaluation is replaced by
\begin{equation}
L_{\mathrm{RU}}^{\mathrm{wc}}[k]
=
L_{\mathrm{FS}}(d_{\mathrm{RU}}(t_k),f_{\mathrm{c}})
+
\hat{X}_{\mathrm{RU}}^{\mathrm{wc}}[k],
\label{eq:Lru_wc}
\end{equation}
and the corresponding conservative large-scale gain is
\begin{equation}
\beta_{\mathrm{RU}}^{\mathrm{wc}}[k]
=
10^{-L_{\mathrm{RU}}^{\mathrm{wc}}[k]/10}.
\label{eq:beta_ru_wc}
\end{equation}

Accordingly, for the $m$th RIS codeword, the communication and positioning metrics are evaluated under the conservative shadowing condition as follows:
\begin{align}
\gamma_m^{\mathrm{rob}}[k]
&\triangleq
\gamma\!\left(t_k,\mathbf{\Theta}_m;\hat{X}_{\mathrm{RU}}^{\mathrm{wc}}[k]\right),
\\
\mathrm{PEB}_m^{\mathrm{rob}}[k]
&\triangleq
\mathrm{PEB}\!\left(t_k,\mathbf{\Theta}_m;\hat{X}_{\mathrm{RU}}^{\mathrm{wc}}[k]\right).
\end{align}
In other words, $\gamma_m^{\mathrm{rob}}[k]$ and $\mathrm{PEB}_m^{\mathrm{rob}}[k]$ are obtained by replacing the nominal RIS–user gain $\beta_{\mathrm{RU}}(t_k)$ in the reflected-channel model and the associated delay-variance expressions with the conservative gain $\beta_{\mathrm{RU}}^{\mathrm{wc}}[k]$. Accordingly, the robust layer developed here focuses on the dominant uncertainty source considered in this work, namely the local large-scale shadowing on the terrestrial RIS–user hop. Other uncertainty sources, such as phase errors, geometry uncertainty, or imperfect knowledge of the direct-link blockage factor, are not explicitly modeled and are left for future investigation.

The resulting robust utility is defined as
\begin{equation}
\mathcal{U}_m^{\mathrm{rob}}[k]
=
\alpha[k]\,
\frac{\gamma_m^{\mathrm{rob}}[k]}{\gamma_{\mathrm{ref}}}
+
\bigl(1-\alpha[k]\bigr)\,
\frac{\mathrm{PEB}_{\mathrm{ref}}}{\mathrm{PEB}_m^{\mathrm{rob}}[k]}.
\label{eq:utility_rob}
\end{equation}

Therefore, the selected RIS configuration is obtained as:
\begin{equation}
m^{\star}[k]
=
\arg\max_{m\in\{1,\ldots,M\}}
\mathcal{U}_m^{\mathrm{rob}}[k],
\label{eq:mstar_rob}
\end{equation}
and
\begin{equation}
\mathbf{\Theta}^{\star}[k]
=
\mathbf{\Theta}_{m^{\star}[k]}.
\label{eq:theta_star_rob}
\end{equation}

This robust layer does not introduce an additional operating mode beyond \eqref{eq:alpha_modes}. Instead, it stabilizes the codebook selection within each blockage-dependent mode by explicitly accounting for the filtered local shadowing condition and its uncertainty.

\subsection{Codebook-Based RIS Configuration}

To maintain low implementation complexity and enable a practically implementable, low-overhead design, we adopt a finite RIS codebook
\begin{equation}
\mathcal{C}=
\left\{
\mathbf{\Theta}_{1},
\mathbf{\Theta}_{2},
\ldots,
\mathbf{\Theta}_{M}
\right\},
\label{eq:codebook}
\end{equation}
where each codeword represents a feasible RIS phase profile selected from the $b$-bit hardware-constrained set.

For the $m$th codeword, the nominal communication and positioning metrics are as follows:
\begin{align}
\gamma_m(t) &\triangleq \gamma\!\left(t,\mathbf{\Theta}_{m}\right), \\
\mathrm{PEB}_m(t) &\triangleq \mathrm{PEB}\!\left(t,\mathbf{\Theta}_{m}\right), \\
\mathcal{U}_m(t) &\triangleq \mathcal{U}\!\left(t,\mathbf{\Theta}_{m}\right).
\end{align}
These nominal quantities are useful for reference and benchmarking. However, the proposed controller performs the final RIS selection according to the robust utility in \eqref{eq:utility_rob}--\eqref{eq:theta_star_rob}, rather than the nominal utility alone.

The codebook in \eqref{eq:codebook} may contain beams designed for different roles; for example:
\begin{itemize}
\item \emph{communication-oriented codewords}: maximizing the coherent combination with the direct path;
\item \emph{positioning-oriented codewords}: strengthening the observability of the reflected-path delay;
\item \emph{balanced codewords}: providing a compromise between the two objectives.
\end{itemize}
The codebook is viewed here as an offline design component. Depending on the implementation, its codewords may be constructed analytically from nominal propagation geometry, selected heuristically to cover communication- and positioning-relevant operating regimes, generated randomly and filtered according to feasibility and performance criteria, or optimized offline over representative scenarios. The emphasis of this work is not on codebook synthesis itself but on robust, low-complexity online selection from a finite feasible codebook under blockage-aware and shadowing-aware operation.

This codebook-based formulation offers two important advantages and is consistent with practical low-overhead RIS control frameworks based on finite reflection-pattern codebooks \cite{An2024}. First, it significantly reduces online optimization complexity, since only $M$ candidate configurations are evaluated per coherence block. Second, it is well suited to practical RIS control, where phase profiles are often precomputed offline and then selected online using low-rate control signaling. Therefore, the finite codebook in \eqref{eq:codebook} provides the implementation structure, while the blockage-aware and shadowing-aware robust utility determines the final codeword choice.

\subsection{Communication- and Positioning-Oriented Benchmarks}

To benchmark the proposed joint design, it is useful to define two single-objective RIS reference configurations corresponding to the communication- and positioning-oriented extremes.

First, neglecting the positioning term, the communication-oriented RIS configuration should maximize the instantaneous SNR in \eqref{eq:gamma}, i.e.,
\begin{equation}
\max_{\mathbf{\Theta}(t)}
\left|
h_{\mathrm{d}}(t)+h_{\mathrm{r}}(t,\mathbf{\Theta})
\right|^{2}.
\label{eq:comm_only}
\end{equation}
For continuous phase shifts, this benchmark corresponds to aligning the reflected path as coherently as possible with the direct path, following the standard communication-oriented RIS beamforming interpretation in the RIS literature \cite{Basar2019,Wu2020a}.

Second, neglecting the communication term, the positioning-oriented RIS configuration should minimize the position error bound, i.e.,
\begin{equation}
\min_{\mathbf{\Theta}(t)}
\mathrm{PEB}(t,\mathbf{\Theta}).
\label{eq:pos_only}
\end{equation}
This benchmark prioritizes reflected-path delay observability and therefore serves as the positioning-oriented counterpart to \eqref{eq:comm_only}.

Together, \eqref{eq:comm_only} and \eqref{eq:pos_only} define two extreme operating points of the communication--positioning design space: the former prioritizes downlink reliability, whereas the latter prioritizes positioning accuracy. In the finite-codebook implementation adopted in this work, these benchmarks are realized by selecting, respectively, the codeword that maximizes the instantaneous SNR or minimizes the PEB over the candidate set in \eqref{eq:codebook}. The proposed joint design generalizes these two extremes by optimizing the utility in \eqref{eq:utility}, while the shadowing-aware robust controller further refines the online codeword selection through \eqref{eq:utility_rob}–\eqref{eq:theta_star_rob}. Hence, the proposed method enables a controlled tradeoff between communication reliability and positioning support under uncertain urban shadowing.

\subsection{Complexity and Online Operation}

At each coherence block, the proposed design first updates the scalar shadowing-state estimate using the recursion introduced in Section II and then evaluates the robust utility of $M$ candidate RIS configurations according to \eqref{eq:utility_rob}. Since the evaluation of the reflected channel for each codeword involves all $N$ RIS elements, the dominant online computational complexity scales as
\begin{equation}
\mathcal{O}(MN),
\end{equation}
whereas the shadowing-state update contributes only a negligible scalar-filter overhead of order $\mathcal{O}(1)$ per block. In practical terms, this means that the controller only needs to score $M$ candidate phase profiles per coherence block, with a per-codeword cost that grows linearly with the number of RIS elements. Thus, if either the codebook size or the RIS size increases, the online cost increases only proportionally without requiring iterative continuous-phase optimization or high-dimensional matrix-based processing. Hence, the proposed shadowing-aware extension preserves the low-complexity structure of codebook-based RIS control and is well suited for real-time online RIS selection. The offline construction of the RIS codebook is not included in this online complexity count.

The resulting online procedure is summarized as follows:
\begin{enumerate}
\item estimate the direct-link blockage factor $\xi_{\mathrm{blk}}[k]$ and determine the corresponding priority weight $\alpha[k]$ from \eqref{eq:alpha_modes};
\item acquire the reflected pilot-power measurement and update the shadowing-state estimate $\hat{X}_{\mathrm{RU}}[k|k]$ and its uncertainty $P_X[k|k]$ using the tracker in Section II;
\item construct the conservative shadowing level $\hat{X}_{\mathrm{RU}}^{\mathrm{wc}}[k]$ according to \eqref{eq:shadow_wc};
\item evaluate $\mathcal{U}_m^{\mathrm{rob}}[k]$ for all $\mathbf{\Theta}_m\in\mathcal{C}$;
\item select $\mathbf{\Theta}^{\star}[k]$ according to \eqref{eq:theta_star_rob}.
\end{enumerate}

Therefore, the proposed RIS controller remains lightweight while still adapting to both the communication--positioning tradeoff induced by urban blockage and the uncertainty caused by spatially correlated RIS--user shadowing.

\section{Results and Evaluation}
\label{sec:results}

In this section, we evaluate the proposed RIS-assisted joint communication–positioning framework under urban NTN conditions. The objective is to quantify how the proposed blockage-aware and shadowing-aware RIS controller improves both communication reliability and two-dimensional positioning accuracy relative to the considered benchmark schemes. In particular, the numerical study examines the effects of blockage severity, user geometry, RIS operating mode, and hardware capability on the received SNR, the PEB, the codeword switching stability, and the joint success probability. Unless otherwise stated, the results are obtained using the simulation parameters summarized in Table~\ref{tab:sim_parameters}.
\begin{table}[!t]
\renewcommand{\arraystretch}{1.05}
\setlength{\tabcolsep}{3pt}
\caption{Main simulation parameters used in the numerical evaluation.}
\label{tab:sim_parameters}
\centering
\footnotesize
\begin{tabular}{p{0.53\columnwidth} p{0.32\columnwidth}}
\hline
\textbf{Parameter} & \textbf{Value} \\
\hline
Carrier frequency, $f_c$ & $2.2~\mathrm{GHz}$ \\
Signal bandwidth, $B$ & $20~\mathrm{MHz}$ \\
Receiver noise figure & $7~\mathrm{dB}$ \\
Satellite EIRP density & $31~\mathrm{dBW/MHz}$ \\
Speed of light, $c$ & $3\times10^{8}~\mathrm{m/s}$ \\
Satellite position, $\mathbf{p}_{\mathrm S}$ & $[350000,0,600000]^{\mathrm T}~\mathrm{m}$ \\
RIS position, $\mathbf{p}_{\mathrm R}$ & $[100,0,20]^{\mathrm T}~\mathrm{m}$ \\
User height, $z_0$ & $1.5~\mathrm{m}$ \\
Atmospheric loss on direct link, $L_{\mathrm{atm,SU}}$ & $1.0~\mathrm{dB}$ \\
Atmospheric loss on sat.--RIS link, $L_{\mathrm{atm,SR}}$ & $0.8~\mathrm{dB}$ \\
RIS size, $N$ & $64$--$1024$ elements \\
Phase resolution, $b$ & $1$--$6$ bits \\
Codebook size, $M$ & $9$ codewords \\
Blockage factor, $\xi_{\mathrm{blk}}$ & $0.01$--$1.0$ \\
RIS--user shadowing std., $\sigma_{\mathrm{RU}}$ & $6$--$8~\mathrm{dB}$ \\
Delay coefficient, $\kappa_d$ & $10^{-2}$ \\
Delay coefficient, $\kappa_r$ & $10^{-2}$ \\
Thresholds, $(\xi_L,\xi_H)$ & $(0.08,0.30)$ \\
Weights, $(\alpha_C,\alpha_B,\alpha_P)$ & $(0.92,0.55,0.15)$ \\
Robustness parameter, $\nu$ & $1.2$--$1.4$ \\
SNR threshold, $\gamma_{\mathrm{th}}$ & $5$--$6.5~\mathrm{dB}$ \\
PEB threshold, $\eta_{\mathrm{th}}$ & $16$--$20~\mathrm{m}$ \\
\hline
\end{tabular}
\end{table}

\subsection{Simulation Setup}

We consider a single LEO satellite, a fixed terrestrial RIS, and a single ground user moving within an urban street region. Unless otherwise stated, the satellite is fixed at a known orbit snapshot for each simulation setup, while the RIS is mounted on a fixed building facade. The user position is sampled over a two-dimensional service region at the known height $z_{0}$ to capture different blockage and geometric conditions. For region-based plots, user locations are sampled across the service area, whereas Monte Carlo results are averaged over repeated realizations. In particular, the communication--positioning tradeoff results in Fig.~\ref{fig:tradeoff_alpha} are averaged over 300 shadowing realizations for each user location on the spatial grid, while the switching-stability results in Fig.~\ref{fig:switching} are averaged over 260 trajectory trials for each shadowing-estimation-noise level.

The direct satellite-to-user and satellite-to-RIS links are modeled using free-space path loss with atmospheric attenuation, whereas the RIS-to-user terrestrial hop is modeled using distance-dependent path loss and spatially correlated shadowing. For each realization, the direct-link blockage factor $\xi_{\mathrm{blk}}$ is generated within the range specified in Table~\ref{tab:sim_parameters}, and the RIS--user shadowing realization is generated according to the spatial model in Section~\ref{sec:spatially_correlated_shadowing}. For the proposed shadowing-aware controller, the RIS--user shadowing state is tracked across coherence blocks using the state-space model and scalar Kalman filter in Section~\ref{sec:shadow_tracking}.

For fairness, the same user position, blockage realization, and shadowing realization are reused across all compared RIS selection methods within each Monte Carlo trial, so that performance differences are attributable only to the RIS control strategy. Unless a figure explicitly studies satellite-geometry variation, the orbit snapshot is kept fixed across trials, while only the user geometry, blockage condition, and shadowing realization are varied.

\subsection{Benchmarks}
The proposed design is compared with the following benchmark schemes.

\subsubsection*{1) Direct-only baseline}
Only the direct satellite-to-user path is used, and the RIS remains inactive. This baseline is used only for communication comparisons since the reduced positioning model requires both the direct-path delay and the RIS-assisted excess-delay measurement.

\subsubsection*{2) Communication-only RIS}
The RIS is configured to maximize the received SNR according to \eqref{eq:comm_only} without explicitly considering positioning performance.

\subsubsection*{3) Positioning-oriented RIS}
The RIS is selected to minimize the PEB without explicitly optimizing communication reliability.

\subsubsection*{4) Joint RIS without shadowing-aware robustness}
The RIS is selected according to the nominal utility in \eqref{eq:utility}, with the blockage-aware weight adaptation in \eqref{eq:alpha_modes}, but without the shadowing-state filter and the conservative margin in \eqref{eq:shadow_wc}.

\subsubsection*{5) Proposed blockage-aware and shadowing-aware joint RIS}
The RIS is selected according to its robust utility in \eqref{eq:utility_rob}, with the communication--positioning priority adapted through \eqref{eq:alpha_modes} and the local RIS--user shadowing uncertainty incorporated through \eqref{eq:shadow_wc}.

\subsubsection*{6) Ideal continuous-phase joint RIS (upper-bound benchmark)}
To assess the gap between the proposed practical codebook-based controller and an idealized RIS design, we also consider a continuous-phase joint benchmark in which the RIS phases are optimized directly according to the nominal utility in \eqref{eq:utility}, without restricting the solution to the finite codebook in \eqref{eq:codebook}. This benchmark removes the finite-codebook constraint and therefore serves as an upper-bound reference for the achievable communication--positioning tradeoff under ideal continuous-phase control.
This benchmark is explicitly plotted in the communication–positioning tradeoff results to quantify the performance gap between the practical finite-codebook controller and the ideal continuous-phase joint design.

\subsection{Evaluation Metrics}

The communication performance is evaluated using the instantaneous SNR, the average SNR, and the outage probability. The positioning performance is evaluated using the average PEB and its spatial variation over the user region. To jointly assess communication and localization resilience, we also use the success probability
\begin{equation}
P_{\mathrm{succ}}=
\Pr\!\left\{
\gamma(t,\mathbf{\Theta})\geq\gamma_{\mathrm{th}},
\ \mathrm{PEB}(t,\mathbf{\Theta})\leq \eta_{\mathrm{th}}
\right\},
\end{equation}
which measures the probability of simultaneously satisfying the communication and positioning requirements.

\subsection{Monte Carlo Evaluation Procedure}

For each simulation realization, the following steps are carried out:
\begin{enumerate}
\item Generate the satellite position, user position, and the corresponding propagation distances.
\item Generate the correlated shadowing realization and the direct-link blockage factor.
\item Construct the direct and RIS-assisted channels according to the channel model in Section~II.
\item For the proposed scheme, update the filtered RIS–user shadowing estimate and construct the conservative shadowing level according to \eqref{eq:shadow_wc}.
\item Determine the operating mode from \eqref{eq:alpha_modes}.
\item For each finite-codebook scheme, evaluate all RIS codewords in the codebook and select the best configuration according to the corresponding selection criterion. For the ideal continuous-phase joint RIS benchmark, directly optimize the nominal utility in (50) over continuous RIS phases without restricting the solution to the finite codebook.
\item Compute the resulting SNR, reduced FIM, PEB, and success event.
\item Repeat the simulation over the prescribed number of realizations and average the resulting performance metrics. The realization counts used in each experiment are specified in the corresponding numerical setup.
\end{enumerate}
\subsection{Numerical Results}

\begin{figure}[!t]
\centering
\includegraphics[width=0.45\textwidth]{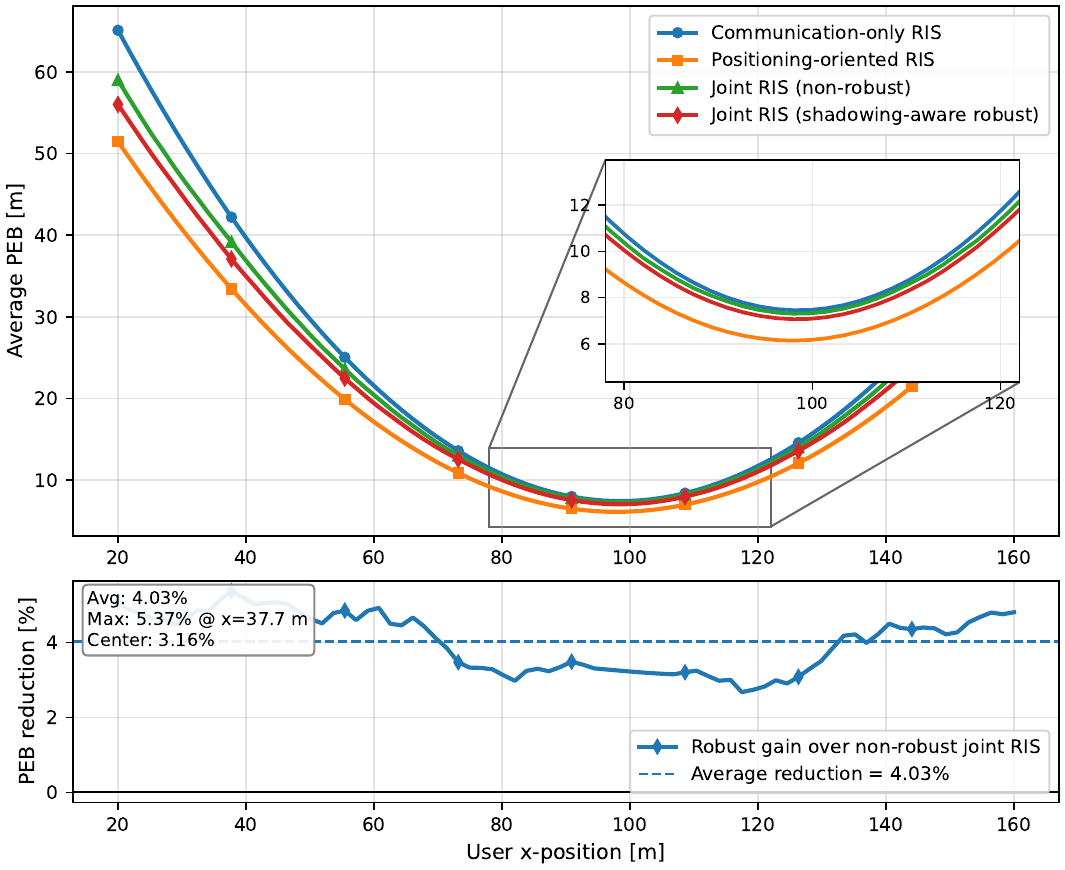}
 \caption{Spatial variation of the average PEB versus the user x-position for different RIS selection strategies. The inset enlarges the central region where the four curves are close. The lower panel reports the percentage PEB reduction of the shadowing-aware robust joint RIS relative to the non-robust joint RIS benchmark.}
\label{fig:peb_vs_position}
\end{figure}

Fig.~\ref{fig:peb_vs_position} shows the spatial variation of the average PEB across the user service region for different RIS selection strategies. A clear U-shaped trend can be observed, indicating that the positioning geometry is more favorable in the middle part of the considered region and becomes weaker toward the edge locations. Although all RIS-active schemes follow the same overall geometry-driven behavior, the inset and the lower panel clarify the differences among the compared designs.

Among the considered schemes, the communication-only RIS design yields the largest PEB because it prioritizes coherent signal enhancement rather than supporting reflected-path positioning. In contrast, the positioning-oriented RIS achieves the smallest PEB, since its codeword selection is dedicated to improving positioning accuracy. The two joint RIS schemes lie between these two extremes, confirming the intended communication–positioning tradeoff of the proposed framework. Compared with the non-robust joint RIS, the proposed shadowing-aware robust joint RIS provides a consistent reduction in average PEB across the user region. In the considered setup, the average PEB reduction is about 4.03\%, with a maximum local reduction of about 5.37\% near $x\approx 37.7$~m. The gain is smaller in the central region, where the competing RIS selections are already close, and becomes more noticeable toward the edge regions, where shadowing uncertainty has a stronger influence on the codeword decision. These results show that the robustness layer yields a modest but practically meaningful positioning benefit while preserving the intended joint-design tradeoff.

\begin{figure}[!t]
\centering
\includegraphics[width=0.44\textwidth]{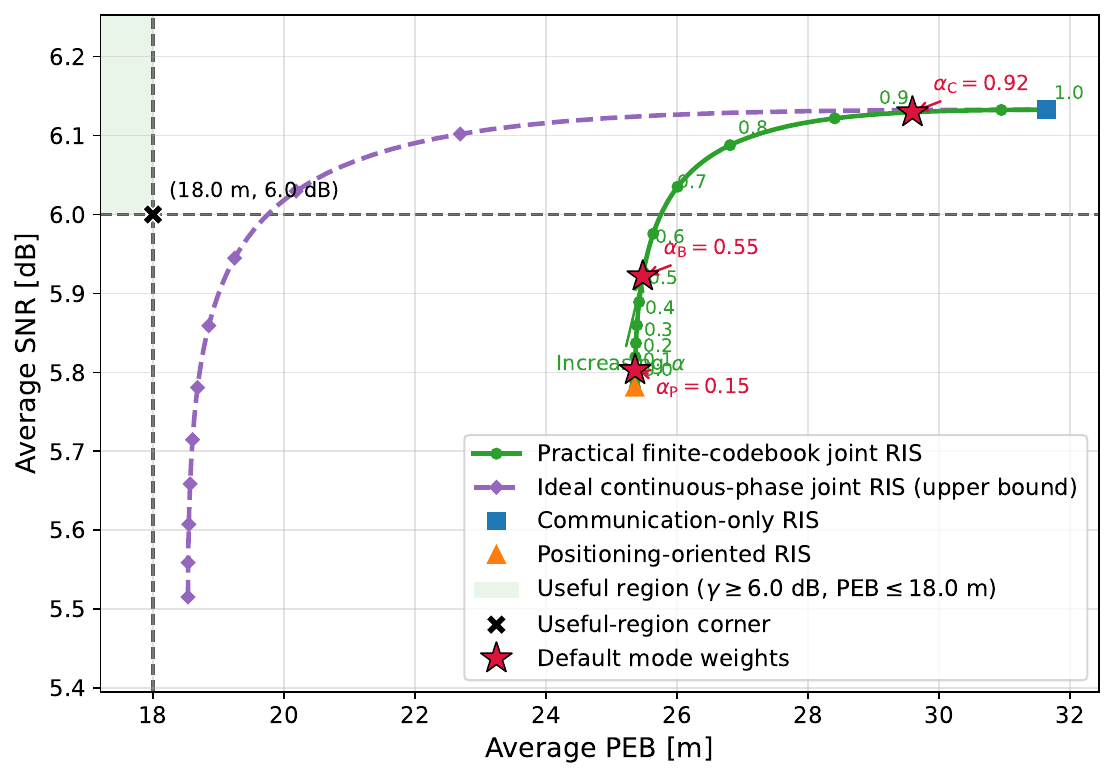}
\caption{Communication--positioning tradeoff in terms of average SNR and average PEB as the weighting parameter $\alpha$ varies from 0 to 1. The green curve corresponds to the practical finite-codebook joint RIS design, while the purple dashed curve shows the ideal continuous-phase joint RIS as an upper-bound benchmark. The communication-only RIS and positioning-oriented RIS indicate the two single-objective extremes. The marked points corresponding to $\alpha_{\mathrm P}$, $\alpha_{\mathrm B}$, and $\alpha_{\mathrm C}$ indicate the default blockage-aware operating points. The dashed threshold lines, shaded useful region, and marked corner point $(18.0~\mathrm{m},\,6.0~\mathrm{dB})$ indicate the representative joint communication--positioning target region.}
\label{fig:tradeoff_alpha}
\end{figure}

Fig.~\ref{fig:tradeoff_alpha} shows the communication--positioning tradeoff achieved by the proposed joint RIS design as the weighting parameter $\alpha$ varies from 0 to 1. Each point on the green curve corresponds to the operating point obtained for a particular value of $\alpha$ under the practical finite-codebook RIS controller. When $\alpha$ is small, the controller gives higher priority to positioning performance, resulting in a smaller average PEB, while the average SNR remains lower. As $\alpha$ increases, the controller gradually shifts its emphasis toward communication quality, which improves the average SNR at the cost of a larger average PEB. Therefore, the practical tradeoff curve confirms the intended behavior of the proposed utility function in \eqref{eq:utility}, namely that it provides a continuous transition between positioning-oriented and communication-oriented RIS operation.

The two single-objective benchmark points further support this interpretation. The positioning-oriented RIS lies near the low-PEB end of the tradeoff, whereas the communication-only RIS lies near the high-SNR end. The proposed joint RIS curve remains between these two extremes, confirming that the controller does not operate as a fixed design but instead provides a tunable balance between communication reliability and positioning accuracy. The marked operating points corresponding to the default mode weights $\alpha_{\mathrm P}$, $\alpha_{\mathrm B}$, and $\alpha_{\mathrm C}$ indicate where the practical blockage-aware controller operates on this tradeoff curve. In addition, the dashed threshold lines and the shaded useful region indicate representative communication and positioning targets, while the marked corner point $(18.0~\mathrm{m},\,6.0~\mathrm{dB})$ explicitly shows the boundary at which both requirements are simultaneously satisfied. Under the present parameter settings, the practical finite-codebook tradeoff curve does not enter the useful region, indicating that the selected requirement pair is relatively stringent for the considered configuration. This result shows that, in the considered operating regime, the main benefit of the proposed design is a controllable communication--positioning tradeoff rather than simultaneous satisfaction of both targets over the entire tradeoff curve. It can also be observed that the variation in PEB is more pronounced than the variation in SNR, suggesting that the weighting parameter $\alpha$ has a stronger influence on positioning performance than on communication quality in this regime.

For reference, Fig.~\ref{fig:tradeoff_alpha} also includes the ideal continuous-phase joint RIS as an upper-bound benchmark. Since this benchmark removes the finite-codebook constraint, it provides a reference for the best achievable communication--positioning tradeoff under idealized RIS phase control. The comparison therefore quantifies the performance loss incurred by practical codebook-based RIS control and phase quantization. It can be observed that the ideal continuous-phase curve lies to the left of, and generally above, the practical finite-codebook curve, meaning that it achieves a better communication--positioning tradeoff overall. In particular, the ideal benchmark yields lower average PEB for a given SNR level, or equivalently higher average SNR for a comparable PEB level, while preserving the same overall tradeoff trend. This behavior confirms that the proposed practical controller follows the intended communication--positioning tradeoff structure, with the remaining gap reflecting the cost of low-complexity implementable RIS operation.

\begin{figure}[!t]
\centering
\includegraphics[width=0.44\textwidth]{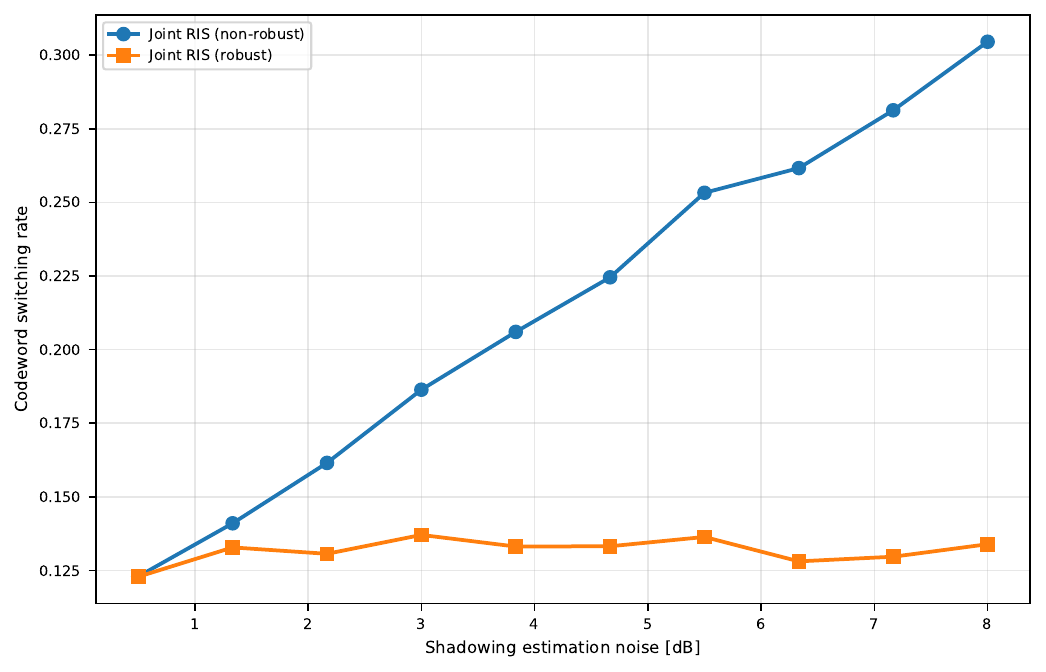}
\caption{Switching stability of the proposed RIS controller versus the shadowing-estimation noise. The robust and non-robust joint RIS designs are compared in terms of codeword switching rate.}
\label{fig:switching}
\end{figure}

Fig.~\ref{fig:switching} shows the switching stability of the proposed RIS controller as a function of the shadowing-estimation noise. As the level of uncertainty increases, the non-robust joint RIS design exhibits a clear rise in the codeword switching rate, indicating that its decisions become increasingly sensitive to noisy shadowing observations. By contrast, the robust controller preserves a substantially lower and nearly flat switching rate across the same range of uncertainty. This result verifies that the proposed shadowing-aware robustness mechanism improves decision stability by filtering local shadowing fluctuations before RIS codeword selection.

Importantly, this improvement in switching stability is achieved without a significant nominal performance penalty. Relative to the non-robust joint RIS, the robust controller changes the average SNR by less than $0.01$~dB on average and reduces the average PEB by about $0.64\%$ under the same realizations across the considered shadowing-noise range. Thus, the robustness layer mainly suppresses spurious codeword changes while preserving nearly the same communication--positioning operating point. 
Therefore, the main benefit demonstrated by Fig.~\ref{fig:switching} is improved codeword-selection stability under shadowing uncertainty, achieved with negligible average SNR loss and only a small change in average PEB.

\begin{figure}[!t]
\centering
\includegraphics[width=0.45\textwidth]{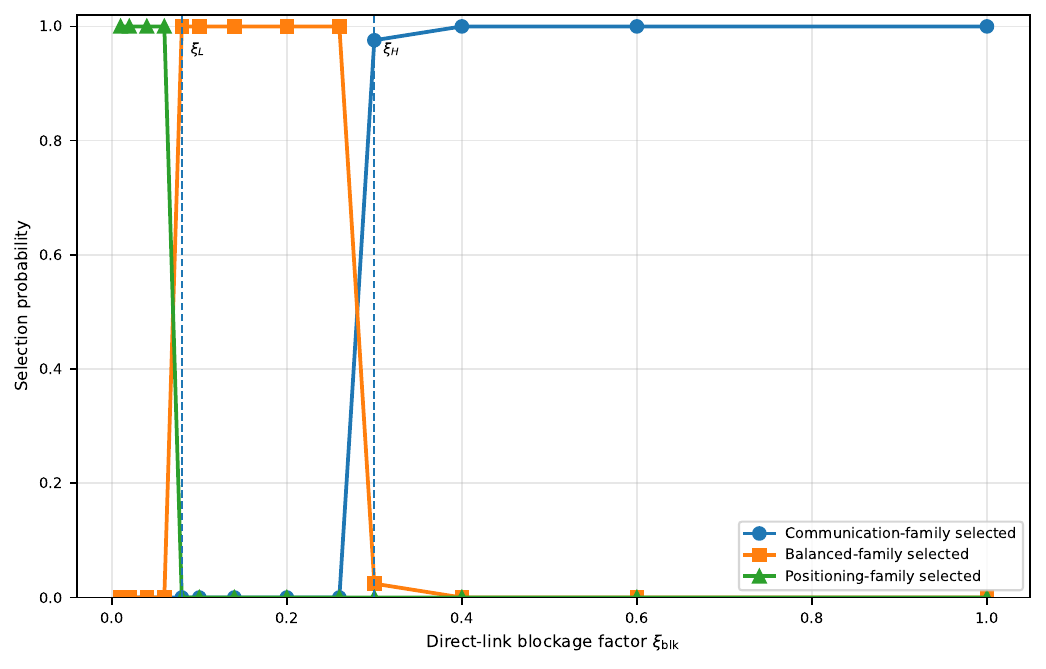}
\caption{Selection probability of the three RIS codeword families versus the direct-link blockage factor $\xi_{\mathrm{blk}}$. The figure confirms the intended three-mode blockage-aware behavior: severe blockage favors positioning-oriented codewords, moderate blockage favors balanced codewords, and weak blockage favors communication-oriented codewords.}
\label{fig:codeword_selection}
\end{figure}

Fig.~\ref{fig:codeword_selection} shows the selection probability of the three RIS codeword families as a function of the direct-link blockage factor $\xi_{\mathrm{blk}}$. A clear three-region behavior is observed, which is consistent with the blockage-aware operating policy in \eqref{eq:alpha_modes}. In the severe-blockage regime, where $\xi_{\mathrm{blk}} \leq \xi_{\mathrm{L}}$, the positioning-oriented family is selected with the highest probability, indicating that the controller shifts its priority toward improving reflected-path positioning support when the direct communication link becomes highly unreliable. In the moderate-blockage regime, where $\xi_{\mathrm{L}} < \xi_{\mathrm{blk}} < \xi_{\mathrm{H}}$, the balanced family becomes dominant, confirming that the RIS controller switches to an intermediate operating mode that jointly supports communication and positioning. In the weak-blockage regime, where $\xi_{\mathrm{blk}} \geq \xi_{\mathrm{H}}$, the communication-oriented family dominates since the direct path is sufficiently reliable, and the RIS is primarily used to enhance the received SNR. Therefore, Fig.~\ref{fig:codeword_selection} quantitatively validates that the proposed blockage-aware controller maps different direct-link conditions to the intended RIS operating mode.

\begin{figure}[!t]
\centering
\includegraphics[width=0.38\textwidth]{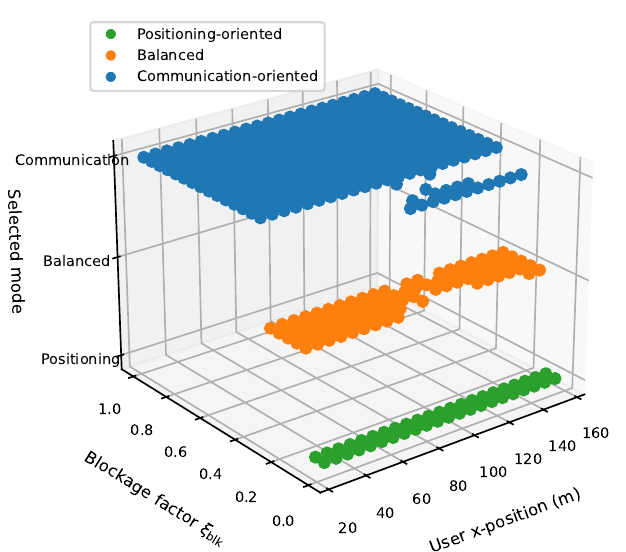}
\caption{3D decision map of the proposed RIS operating mode as a function of the user x-position and the direct-link blockage factor $\xi_{\mathrm{blk}}$. The three decision regions correspond to positioning-oriented, balanced, and communication-oriented RIS operation.}
\label{fig:decision_map_3d}
\end{figure}

Fig.~\ref{fig:decision_map_3d} provides a spatial visualization of the blockage-aware operating-mode selection of the proposed RIS controller as a function of user position and the direct-link blockage factor $\xi_{\mathrm{blk}}$. The three decision regions are consistent with the policy in \eqref{eq:alpha_modes}: severe blockage favors the positioning-oriented mode, moderate blockage favors the balanced mode, and weak blockage favors the communication-oriented mode. Unlike Fig.~\ref{fig:codeword_selection}, which quantifies the mode-selection tendency versus blockage, this figure highlights that the transition boundaries are also influenced by the underlying user geometry. Hence, Fig.~\ref{fig:decision_map_3d} serves mainly as an illustrative spatial complement to the quantitative validation in Fig.~\ref{fig:codeword_selection}.

\begin{figure}[!t]
\centering
\includegraphics[width=0.45\textwidth]{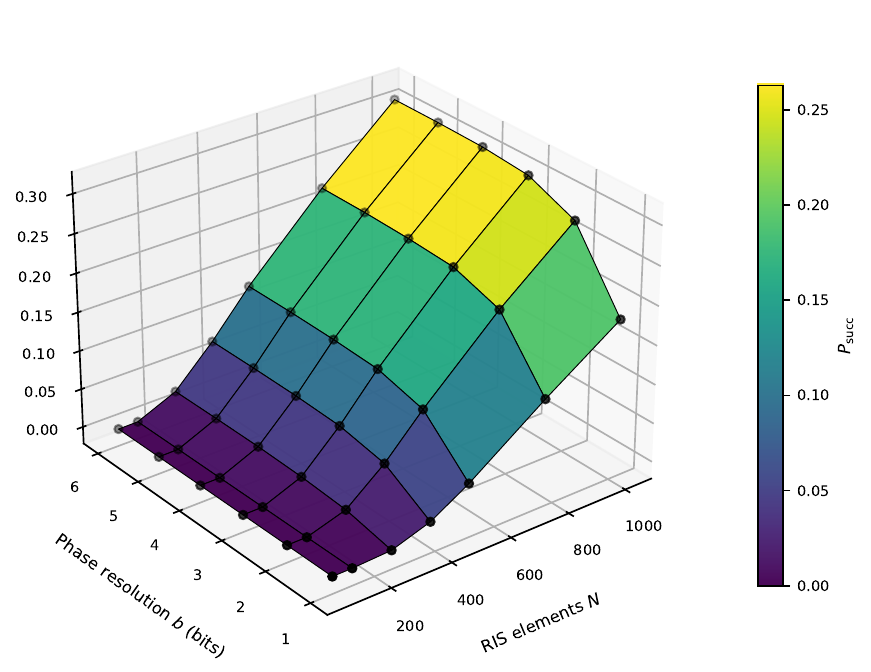}
\caption{3D surface of the joint success probability $P_{\mathrm{succ}}$ versus the RIS size $N$ and the phase-shifter resolution $b$. The success probability improves with both $N$ and $b$, while a diminishing-return trend at higher hardware complexity reveals a practical knee region for RIS design.}
\label{fig:psucc_3d}
\end{figure}

Fig.~\ref{fig:psucc_3d} shows the joint success probability $P_{\mathrm{succ}}$ as a function of the RIS size $N$ and the phase-shifter resolution $b$. It can be observed that the success probability increases as either the number of RIS elements or the phase resolution becomes larger, confirming that both the RIS aperture and phase-control accuracy improve the joint communication–positioning performance. Compared with the previous setting, the impact of the phase resolution is more visible, especially in the medium and large-RIS regimes, where quantization loss becomes a relevant limiting factor. At the same time, the surface reveals a clear diminishing-return behavior: once $N$ and $b$ reach moderate values, the additional increase in $P_{\mathrm{succ}}$ becomes progressively smaller. This trend provides useful hardware-design guidance since it indicates the existence of a practical knee region beyond which further increases in RIS size or phase resolution offer only marginal system-level benefits. In the considered operating regime, the surface also shows that increasing $N$ produces a stronger overall improvement than increasing $b$, implying that RIS aperture should be prioritized before moving to very high phase-shifter resolution. Therefore, Fig.~\ref{fig:psucc_3d} not only validates the expected dependence of $P_{\mathrm{succ}}$ on both $N$ and $b$, but also identifies the range in which moderate hardware complexity already captures most of the achievable joint communication–positioning gain.

\begin{figure}[!t]
\centering
\includegraphics[width=0.5\textwidth]{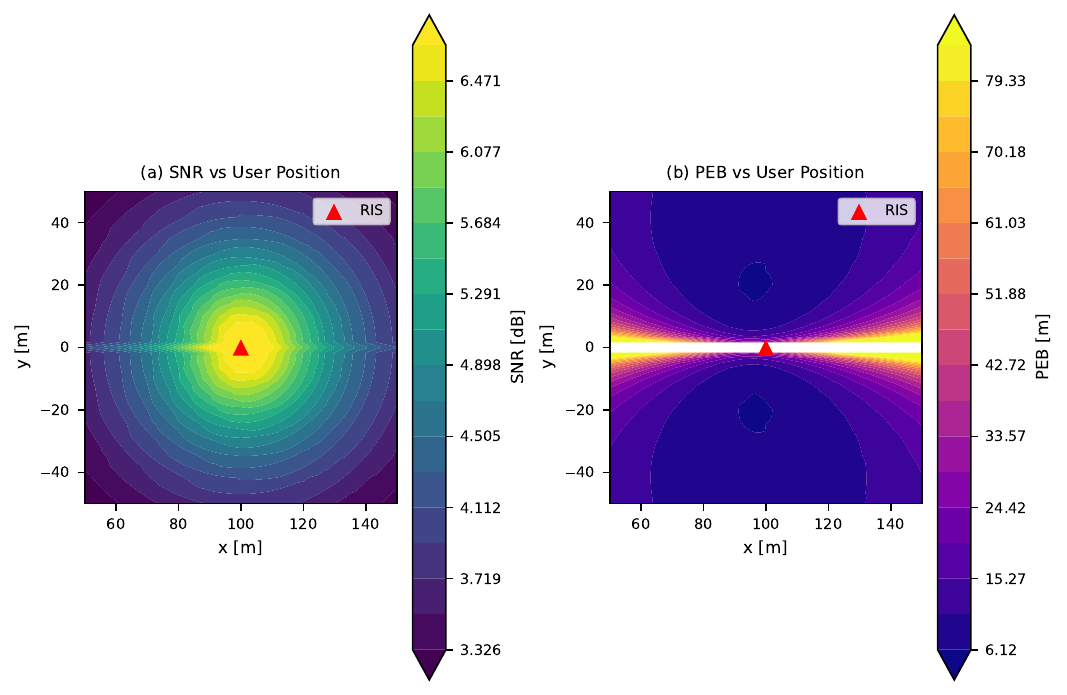}
\caption{Spatial validation of the proposed RIS-assisted framework. (a) Spatial variation of the received SNR over the considered two-dimensional user region. (b) Spatial variation of the PEB over the same region.}
\label{fig:spatial_validation}
\end{figure}

Fig.~\ref{fig:spatial_validation}(a) shows the spatial variation of the received SNR over the considered two-dimensional user region. As expected, the SNR is highest near the RIS and decreases as the user moves farther away. This behavior is consistent with \eqref{eq:gamma}, since the RIS-assisted contribution depends strongly on the RIS--user link gain and therefore benefits from a shorter RIS--user distance. Hence, Fig.~\ref{fig:spatial_validation}(a) confirms that the communication benefit of the RIS is strongly geometry-dependent.

Fig.~\ref{fig:spatial_validation}(b) shows the corresponding spatial variation of the PEB. Smaller PEB values indicate better positioning accuracy, whereas larger values indicate poorer performance. Unlike the SNR map, the PEB is not determined solely by signal strength. According to \eqref{eq:FIM_reduced} and \eqref{eq:PEB}, it depends on both the quality of the delay observations and the geometric diversity of the direct-delay and RIS-assisted excess-delay measurements. Consequently, the region that is favorable for positioning does not exactly coincide with the region of strongest received SNR.

Taken together, Fig.~\ref{fig:spatial_validation}(a) and Fig.~\ref{fig:spatial_validation}(b) reveal a clear spatial mismatch between communication-favorable and positioning-favorable regions. This directly motivates the joint utility in \eqref{eq:utility}: optimizing the RIS only for communication would bias the design toward stronger RIS--user links, whereas optimizing it only for positioning would bias the design toward more favorable delay-domain geometry. Since these objectives are not spatially aligned, practical RIS control must balance both rather than optimize either one in isolation.

For convenience, Table~\ref{tab:summary_default} summarizes the main communication, positioning, and resilience metrics under a representative default fixed setup. Table~\ref{tab:summary_switching} separately summarizes the switching-stability comparison between the robust and non-robust joint RIS designs under the shadowing-estimation-noise sweep in Fig.~\ref{fig:switching}. Because the two tables correspond to different experiments, their entries should be interpreted separately.

\begin{table}[!t]
\renewcommand{\arraystretch}{1.05}
\setlength{\tabcolsep}{4pt}
\caption{Representative quantitative summary under the default fixed setup.}
\label{tab:summary_default}
\centering
\footnotesize
\begin{tabular}{lccc}
\hline
\textbf{Scheme} & \textbf{Avg. SNR [dB]} & \textbf{Avg. PEB [m]} & \textbf{$P_{\mathrm{succ}}$} \\
\hline
Communication-only RIS & 6.133 & 31.637 & 0.360 \\
Positioning-oriented RIS & 5.781 & 25.355 & 0.308 \\
Joint RIS ($\alpha_{\mathrm P}=0.15$) & 5.803 & 25.358 & 0.308 \\
Joint RIS ($\alpha_{\mathrm B}=0.55$) & 5.922 & 25.476 & 0.306 \\
Joint RIS ($\alpha_{\mathrm C}=0.92$) & 6.130 & 29.593 & 0.360 \\
\hline
\end{tabular}

\vspace{1mm}
\raggedright
\footnotesize
Here, $P_{\mathrm{succ}}$ is evaluated using $\gamma_{\mathrm{th}}=6.0$~dB and $\eta_{\mathrm{th}}=18.0$~m.
\end{table}

\begin{table}[!t]
\renewcommand{\arraystretch}{1.02}
\setlength{\tabcolsep}{3pt}
\caption{Switching summary for Fig.~\ref{fig:switching}.}
\label{tab:summary_switching}
\centering
\footnotesize
\begin{tabular}{lccc}
\hline
\textbf{Design} & \textbf{Switch} & \textbf{$\Delta$SNR} & \textbf{$\Delta$PEB} \\
 & \textbf{rate} & \textbf{[dB]} & \textbf{[\%]} \\
\hline
Non-robust & 0.214 & 0 & 0 \\
Robust     & 0.132 & $<0.01$ & $-0.64$ \\
\hline
\end{tabular}

\vspace{1mm}
\raggedright
\footnotesize
The $\Delta$SNR and $\Delta$PEB entries are reported relative to the non-robust design.
\end{table}

\section{Conclusion}
This paper proposed a blockage-aware and shadowing-aware RIS-assisted framework for joint communication and positioning in urban NTNs. A terrestrial RIS was used to support a blockage-sensitive LEO satellite downlink by creating an additional reflected path that improves both received signal quality and user-position observability. A reduced two-dimensional delay-domain positioning model was developed and integrated with a joint utility based on the received SNR and the PEB.

Based on this formulation, a blockage-aware three-mode RIS operating policy was introduced to adapt the controller among communication-oriented, balanced, and positioning-oriented modes according to the direct-link condition. A shadowing-aware robust codebook-based selection mechanism was also developed by tracking the RIS--user shadowing state and incorporating its uncertainty into the online RIS decision process, while preserving low online complexity.

The numerical results showed that the proposed framework achieves a controllable communication--positioning tradeoff, improves positioning performance relative to communication-only RIS selection while maintaining competitive SNR performance, and provides more stable codeword selection under noisy shadowing observations. The results also verified the intended three-mode blockage-aware operation and showed that the joint success probability improves with RIS size and phase resolution, with diminishing returns at high hardware complexity.

These results highlight the importance of designing RIS control in urban NTN jointly for communication reliability and positioning observability rather than for signal enhancement alone. Future work includes extensions to multi-user and multi-satellite scenarios, more detailed wideband or near-field RIS models, and learning-based online adaptation under imperfect channel and blockage estimation.

\bibliographystyle{IEEEtran}
\bibliography{Ref}

\end{document}